\documentclass{sigchi}

\toappear{\scriptsize
Permission to make digital or hard copies of all or part of this work for
personal or classroom use is granted without fee provided that copies are not
made or distributed for profit or commercial advantage and that copies bear
this notice and the full citation on the first page. Copyrights for components
of this work owned by others than the author(s) must be honored. Abstracting
with credit is permitted. To copy otherwise, or republish, to post on servers
or to redistribute to lists, requires prior specific permission and/or a fee.
Request permissions from permissions@acm.org. \\
{\emph{UIST '20, October 20--23, 2020, Virtual Event, USA} } \\
\copyright~2020 Copyright is held by the owner/author(s).
Publication rights licensed to ACM. \\
ACM ISBN 978-1-4503-7514-6/20/10\ ...\$15.00.\\
http://dx.doi.org/10.1145/3379337.3415890}

\ifdefined\SkylineProceedings
\else
\fi

\usepackage{balance}       
\usepackage{graphics}      
\usepackage[T1]{fontenc}   
\usepackage{txfonts}
\usepackage{mathptmx}
\usepackage[pdflang={en-US},pdftex]{hyperref}
\usepackage{color}
\usepackage{booktabs}
\usepackage{textcomp}

\usepackage{microtype}        
\usepackage{ccicons}          
\usepackage[utf8]{inputenc}

\usepackage[inline]{enumitem}
\usepackage{graphicx}
\usepackage[tight]{subfigure}
\usepackage{cite}
\usepackage{amsmath}
\usepackage{amssymb}
\usepackage{overpic}
\usepackage{inconsolata}
\usepackage{booktabs}
\usepackage{tabularx}

\usepackage{pgfplots,pgfplotstable}
\usepgfplotslibrary{groupplots}

\definecolor{dbpalette0}{HTML}{194D7C}
\definecolor{dbpalette2}{HTML}{DCF2FF}

\definecolor{lgray}{HTML}{BBBBBB}

\pgfplotsset{compat=1.13,
  /pgfplots/ybar legend/.style={
    /pgfplots/legend image code/.code={%
       \draw[##1,/tikz/.cd,yshift=-0.25em]
        (0cm,0cm) rectangle (3pt,0.8em);},
   },
}

\usepackage{tikz}
\newcommand*\captionlabel[1]{%
  \tikz[baseline=(char.base)]{%
    \node[shape=circle,fill=white,draw=black,inner sep=2pt] (char) {%
      \textcolor{black}{\footnotesize\textsf{#1}}};
}}

\def\plaintitle{Skyline: Interactive In-Editor Computational Performance
  Profiling for Deep Neural Network Training}
\def\plainauthor{Geoffrey X. Yu, Tovi Grossman, Gennady Pekhimenko}
\def\plainkeywords{Skyline; interactive performance profiling; debugging;
  visualization; deep neural networks; machine learning.}

\makeatletter
\def\url@leostyle{%
  \@ifundefined{selectfont}{
    \def\UrlFont{\sf}
  }{
    \def\UrlFont{\small\bf\ttfamily}
  }}
\makeatother
\def\pprw{8.5in}
\def\pprh{11in}

\usepackage{adjustbox}
\usepackage{array}
\newcolumntype{R}[2]{%
    >{\adjustbox{angle=#1,lap=\width-(#2)}\bgroup}%
    l%
    <{\egroup}%
}
\newcommand*\rot[2]{\multicolumn{1}{R{#1}{#2}}}%
\newcommand{\Sup}{\checkmark}
\newcommand{\NSup}{\color{lgray}--}

\definecolor{linkColor}{RGB}{6,125,233}
\hypersetup{%
  pdftitle={\plaintitle},
  pdfauthor={\plainauthor},
  pdfkeywords={\plainkeywords},
  pdfdisplaydoctitle=true, 
  bookmarksnumbered,
  pdfstartview={FitH},
  colorlinks,
  citecolor=black,
  filecolor=black,
  linkcolor=black,
  urlcolor=linkColor,
  breaklinks=true,
  hypertexnames=false
}

\newcommand\thetool{\textsc{Skyline}}
\newcommand\namedparagraph[1]{\textbf{#1}}
\newcommand{\MemAvgError}{0.19\%}
\newcommand{\MemMaxError}{1.1\%}
\newcommand{\ThptAvgError}{3.7\%}
\newcommand{\ThptMaxError}{10.\%}
\newcommand{\MemMaxStdDev}{0.00949~GiB}
\newcommand{\ThptMaxStdDev}{1.77~samples/s}

\begin{document}

\title{\plaintitle}

\numberofauthors{3}
\author{%
  Geoffrey X. Yu\textsuperscript{*\ddag{}},
  Tovi Grossman\textsuperscript{*},
  Gennady Pekhimenko\textsuperscript{*\ddag{}}\\
  \begin{tabular}[t]{p\auwidth}
    \centering\baselineskip 13pt
    \affaddr{\textsuperscript{*}University of Toronto}\\
    \affaddr{Toronto, Ontario, Canada}
  \end{tabular}
  \hskip 1pt
  \begin{tabular}[t]{p\auwidth}
    \centering\baselineskip 13pt
    \affaddr{\textsuperscript{\ddag}Vector Institute}\\
    \affaddr{Toronto, Ontario, Canada}
  \end{tabular}\\
  \begin{tabular}[t]{c}
    \email{gxyu@cs.toronto.edu,
      tovi@dgp.toronto.edu,
      pekhimenko@cs.toronto.edu}
  \end{tabular}
}

\teaser{%
  \centering
  \subfigure{%
    \begin{overpic}[width=\textwidth]{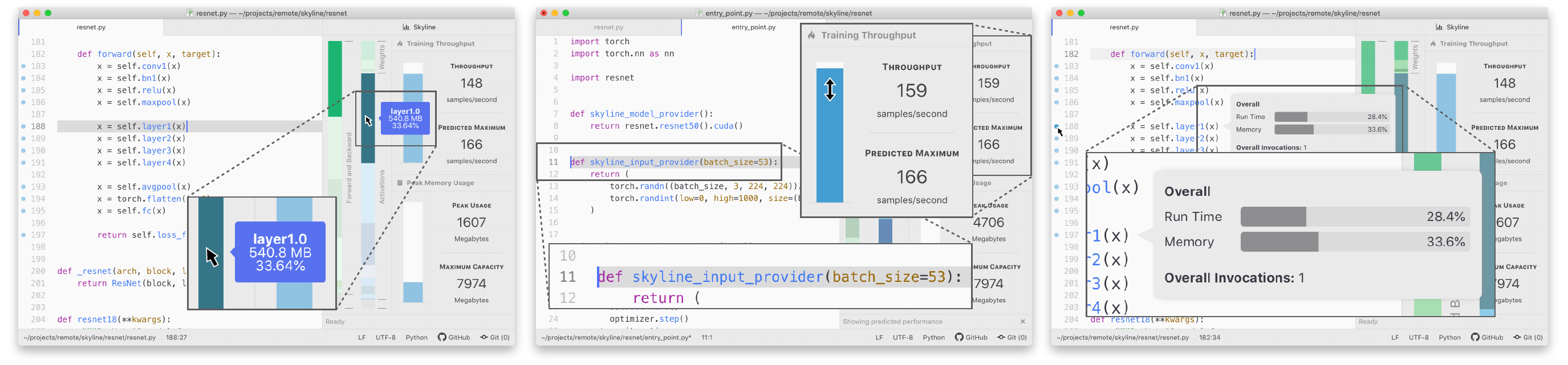}
      \put(2,19.5){\captionlabel{A}}
      \put(35,19.5){\captionlabel{B}}
      \put(68,19.5){\captionlabel{C}}
    \end{overpic}
    \label{fig:overview-hover}
    \addtocounter{subfigure}{1}
    \label{fig:overview-predict}
    \addtocounter{subfigure}{1}
    \label{fig:overview-inline}
    \addtocounter{subfigure}{1}
  }
  \caption{An overview of \thetool{} and its features.
    \subref{fig:overview-hover} Hovering over the visualizations highlights the
    relevant line(s) of associated code and reveals detailed performance
    information.
    \subref{fig:overview-predict} Manipulating the throughput and memory usage
    bar charts results in predictive changes to the batch size that help
    achieve the desired performance metrics.
    \subref{fig:overview-inline} Hovering over markers for specific lines of
    code reveals performance details for that line of code.
  }\label{fig:overview}
  \vspace{-0.5em}
}

\maketitle

\begin{abstract}
Training a state-of-the-art deep neural network (DNN) is a
computationally-expensive and time-consuming process, which incentivizes deep
learning developers to debug their DNNs for computational performance.
However, effectively performing this debugging requires intimate knowledge
about the underlying software and hardware systems---something that the typical
deep learning developer may not have.
To help bridge this gap, we present \thetool{}: a new interactive tool for DNN
training that supports in-editor computational performance profiling,
visualization, and debugging.
\thetool{}'s key contribution is that it leverages special computational
properties of DNN training to provide
\begin{enumerate*}[label=(\roman*)]
  \item interactive performance predictions and visualizations, and
  \item directly manipulatable visualizations that, when dragged, mutate the
    batch size in the code.
\end{enumerate*}
As an in-editor tool, \thetool{} allows users to leverage these diagnostic
features to debug the performance of their DNNs during development.
An exploratory qualitative user study of \thetool{} produced promising results;
all the participants found \thetool{} to be useful and easy to use.
\end{abstract}




\section{Introduction}
In recent years, deep neural networks (DNNs\footnote{Note that in this paper,
  we also refer to DNNs as \emph{models} and use the two terms
  interchangeably.}) have led to breakthroughs in classical machine learning
tasks such as image
classification~\cite{resnet-he16,vgg-simonyan15,alexnet-krizhevsky12,densenet-huang17},
object detection~\cite{maskrcnn-he17}, machine
translation~\cite{nmt-sutskever14,attention-vaswani17}, speech
recognition~\cite{deepspeech2-amodei15}, and
recommendations~\cite{deeprec-covington16}.
As a result, there has been an exciting effort by researchers and practitioners
to apply DNNs in practice.

However, a central concern when using DNNs in practice is that they can be
computationally-expensive, memory-intensive, and time-consuming to
\emph{train}---ranging from hours to
\emph{weeks}~\cite{mlperf,dawn-coleman17,tbd-zhu18,bert-devlin19,mlenergy-strubell19}.
Not only can this process cost users time, but it can also translate to
substantial monetary costs (e.g., cloud compute costs) and, indirectly,
non-trivial amounts of carbon emissions.
For example, Strubell et al. found that training BERT~\cite{bert-devlin19}, a
state-of-the-art DNN used for language modeling, can result in carbon emissions
roughly equivalent to a trans-American flight and would incur over \$12,000 in
cloud compute costs~\cite{mlenergy-strubell19}.
To make matters worse, the computational cost of DNN training is only projected
to continue increasing, doubling every 3.4 months~\cite{aicompute-openai18}.

As a result, these costs incentivize deep learning developers to debug and tune
their DNNs for \emph{computational performance}.\footnote{In this paper, when
  we refer to a DNN's performance we mean its \emph{computational} training
  performance---\emph{not} its prediction accuracy.}
This tuning involves experimenting with different DNN designs, as well as
adjusting parameters such as the \emph{batch size}: the number of data samples
processed during a single training step.
For example, prior work found that larger batch sizes could lead to a higher
training throughput for certain models~\cite{mlperf,tbd-zhu18,dawn-coleman17},
which helped motivate later efforts to enable training with larger batch sizes
on the same hardware platforms~\cite{echo-zheng20,checkmate-jain20}.

Today, effectively performing this kind of debugging and tuning requires
intimate knowledge of and visibility through the \emph{entire} software and
hardware stack---something the typical deep learning developer may not have.
This is because deep learning developers primarily work at the top of this
stack, implementing their models using high-level software frameworks such as
PyTorch~\cite{pytorch-paszke19} or TensorFlow~\cite{tensorflow-abadi16}.
By design, these frameworks abstract away the underlying hardware (e.g.,
graphics processing units (GPUs)~\cite{turing-nvidia} or tensor processing
units (TPUs)~\cite{tpu-jouppi17}) and software systems used during training,
which are critical for tuning, understanding, and debugging performance.
This abstraction results in a \emph{loss of information}, where developers may
face challenges connecting the high-level code they write to the performance
characteristics of their models (e.g., training throughput and memory usage).

Unfortunately, much of the existing work on machine learning performance
debugging has not addressed computational performance, and instead has focused
on \emph{model quality} (i.e. helping improve a model's prediction
accuracy)~\cite{modeltracker-amershi15,visinspect-krause16,seq2seqvis-strobelt19,gestalt-patel10,activis-kahng18}.
To help bridge this gap, we make the key observation that DNN training has a
set of favorable \emph{properties} that enable the use of interactive features
that can help with performance understanding and debugging.
DNN training
\begin{enumerate*}[label=(\roman*)]
  \item consists of \emph{short repetitive iterations}, enabling rapid
    profiling;
  \item has \emph{predictable} performance with respect to batch size changes,
    enabling performance suggestions; and
  \item is implemented using \emph{structured software frameworks}, enabling
    the linking of performance details to specific lines of code.
\end{enumerate*}

In this work, we leverage these properties to develop \thetool{}: a new
interactive computational performance profiling and debugging tool for DNN
training.
\thetool{} visualizes domain-specific performance metrics \emph{in the editor}
where developers write their code, and also links these visualizations to their
associated lines of code (Figure~\ref{fig:overview}\subref{fig:overview-hover}
and \ref{fig:overview}\subref{fig:overview-inline}).
Some of these features are inspired by prior work on in-editor
tooling~\cite{gestalt-patel10,perfhat-cito19,insitu-hoffswell18,insitucode-beck13}
and live programming environments~\cite{theseus-lieber14,omnicode-kang17}.
\thetool{} extends these ideas to DNN training performance and introduces
directly manipulatable performance visualizations that mutate the batch size in
the code when dragged (Figure~\ref{fig:overview}\subref{fig:overview-predict}).
As a \emph{computational} performance profiling tool, \thetool{} does not make
predictions about a model's accuracy (see our
\hyperref[sec:discussion]{Discussion} section) nor does it provide model
accuracy diagnostics, which has been explored in prior
work~\cite{visinspect-krause16,seq2seqvis-strobelt19,activis-kahng18}.
However, we see \thetool{} as a step toward unified deep learning development
tools that include diagnostics and predictions for both performance and model
accuracy.

Finally, as an initial evaluation, we conducted an exploratory qualitative user
study that examined \thetool{}'s diagnostic capabilities.
In our study, seven deep learning researchers
\begin{enumerate*}[label=(\roman*)]
  \item used \thetool{} to carry out five performance investigation tasks
    across three state-of-the-art DNNs: GNMT~\cite{gnmt-wu16}, the
    Transformer~\cite{attention-vaswani17}, and ResNet-50~\cite{resnet-he16},
    and
  \item completed a qualitative questionnaire about the usefulness and
    ease of use of \thetool{} and its features.
\end{enumerate*}
Our study results were promising, with all participants either agreeing or
strongly agreeing that \thetool{} is useful and easy to use---paving the way
for a more comprehensive future study about how \thetool{} might help influence
model design.

In summary, the contributions of this work are:
\begin{itemize}[noitemsep]
  \item Empirically-derived analytical performance models that predict a DNN's
    training throughput and memory usage given a batch size.
  \item Directly manipulatable visualizations that, when dragged, use these
    performance models to suggest batch sizes in the code that may achieve a
    desired training throughput or memory usage.
  \item The design, implementation, and exploratory qualitative evaluation of
    \thetool{}: an interactive in-editor computational performance profiling
    tool for DNN training.
\end{itemize}

\section{Related Work}
\thetool{} builds upon bodies of prior work on
\begin{enumerate*}[label=(\roman*)]
  \item interactive model quality debugging
  tools~\cite{modeltracker-amershi15,visinspect-krause16,seq2seqvis-strobelt19,activis-kahng18,gestalt-patel10},
  \item integrated development environment (IDE)-based performance
    tools~\cite{perfhat-cito19,insitu-hoffswell18,insitucode-beck13},
  \item live programming environments for code
    understanding~\cite{pythontutor-guo13,theseus-lieber14,omnicode-kang17},
    and
  \item other similar commercial tools~\cite{tensorflowprofiler,nvprof,pyprof,dlprof}.
\end{enumerate*}
The key fundamental differences between \thetool{} and these prior works are
that \thetool{}
\begin{enumerate*}[label=(\roman*)]
  \item focuses on DNN computational performance understanding and debugging,
    an increasingly important problem as we described previously; and
  \item leverages the special properties of DNN training computation to
    implement new interactive and manipulatable visualizations linked to the
    batch size in the code.
\end{enumerate*}

\subsection{Machine Learning Accuracy Debugging Tools}
Much of the prior work on machine learning performance debugging focuses on
performance from the perspective of \emph{model
  quality}~\cite{modeltracker-amershi15,visinspect-krause16,seq2seqvis-strobelt19,activis-kahng18,gestalt-patel10}.
ModelTracker~\cite{modeltracker-amershi15}, a tool for general supervised
learning models, helps users investigate a model's prediction accuracy on a
dataset both for individual data points and in aggregate.
Prospector~\cite{visinspect-krause16}, ActiVis~\cite{activis-kahng18} and
Seq2Seq-Vis~\cite{seq2seqvis-strobelt19} address model interpretability:
understanding why models make certain predictions.
Gestalt~\cite{gestalt-patel10}, on the other hand, is a development environment
for machine learning---similar to \thetool{} in spirit. However, Gestalt
focuses on the entire machine learning pipeline: loading and processing data,
implementing models, analyzing predictions made by models, and fixing
correctness bugs in the pipeline.

In contrast with all these works, \thetool{} tackles the \emph{computational}
performance aspect of machine learning, with a focus on DNN training.
\thetool{} aims to help users understand and debug the execution performance of
their models during training---an increasingly important problem as DNNs have
become more widely used in recent years.

\subsection{IDE-Based Performance Debugging}
Prior work has also explored the idea of surfacing performance information
inside a text editor (e.g., an IDE)~\cite{perfhat-cito19,insitucode-beck13},
including placing visualizations in situ with the
code~\cite{insitu-hoffswell18}. For example,
PerformanceHat~\cite{perfhat-cito19} helps software engineers debug performance
bottlenecks in general-purpose Java code by surfacing production monitoring
information inside the IDE. \thetool{}'s approach is inspired by these systems,
but is complementary as \thetool{} specifically addresses DNN training
performance. Moreover, \thetool{} leverages the special properties of DNN
training computation to extend these ideas by introducing manipulatable
visualizations linked to the batch size in the code.

\subsection{Live Programming Environments}
\thetool{}'s ability to provide live, up-to-date performance information is
inspired by prior work on live programming
environments~\cite{omnicode-kang17,theseus-lieber14,pythontutor-guo13}.
Theseus~\cite{theseus-lieber14} and Omnicode~\cite{omnicode-kang17} visualize
line-by-line execution results of JavaScript and Python code respectively.
These systems help programming novices understand the execution of their code
and develop mental models about programming. \thetool{} is different because it
takes this idea in the direction of \emph{computational performance} for DNN
training. Through interactive visualizations, \thetool{} aims to help deep
learning users (who may lack systems-level expertise) understand, debug, and
develop mental models about the computational performance of their DNNs.

\subsection{Existing DNN Profiling Tools}
A few existing tools also help with performance debugging. The TensorFlow (TF)
Profiler~\cite{tensorflowprofiler} is similar in spirit to \thetool{}, but is
instead a standalone tool that lacks \thetool{}'s interactive performance
predictions. DLProf~\cite{dlprof}, PyProf~\cite{pyprof}, and PyTorch's built-in
profiler~\cite{pytorch-paszke19} are primarily command-line based DNN
performance profilers, and nvprof~\cite{nvprof} is a general-purpose GPU
performance profiler. Table~\ref{tbl:feature-compare} shows a feature
comparison between \thetool{} and these existing tools. In a nutshell,
\thetool{} is fundamentally different from these existing tools because it
surfaces high-level domain-specific performance information \emph{in-editor}
and has manipulatable visualizations linked to the batch size in the code.

\begin{table}[t]
  \centering
  \footnotesize
  \begin{tabularx}{\columnwidth}{Xllllll}
    \toprule
    \textbf{Feature / Functionality} &
    \rot{45}{1em}{nvprof~\cite{nvprof}} &
    \rot{45}{1em}{DLProf~\cite{dlprof}} &
    \rot{45}{1em}{TF Profiler~\cite{tensorflowprofiler}} &
    \rot{45}{1em}{PyProf~\cite{pyprof}} &
    \rot{45}{1em}{PyTorch~\cite{pytorch-paszke19}} &
    \rot{45}{3em}{\textbf{\thetool{}}} \\
    \midrule
    Shows GPU kernel info & \Sup & \Sup & \Sup & \Sup & \NSup & \NSup \\
    Shows thpt. / iter. run time & \Sup & \Sup & \Sup & \Sup & \Sup & \Sup \\
    Iter. run time breakdown & \NSup & \Sup & \Sup & \Sup & \Sup & \Sup \\
    Shows total memory usage & \NSup & \NSup & P & \NSup & \Sup & \Sup \\
    Memory usage breakdown & \NSup & \NSup & P & \NSup & \Sup & \Sup \\
    Hierarchical breakdowns & \NSup & \NSup & \NSup & \NSup & \NSup & \Sup \\
    In-editor profiling & \NSup & \NSup & \NSup & \NSup & \NSup & \Sup \\
    Code-linked visualizations & \NSup & \NSup & \NSup
      & \NSup & \NSup & \Sup \\
    Interactive thpt. and mem. vs. batch size predictions & \NSup
      & \NSup & \NSup & \NSup & \NSup & \Sup \\
    \bottomrule
  \end{tabularx}
  \caption{A feature comparison between \thetool{} and existing DNN training
    profiling tools, as of July 2020. P = Planned support in the future. See
    the \hyperref[sec:skyline]{\thetool{}} section for a description of these
    features.}
  \label{tbl:feature-compare}
  \vspace{-1em}
\end{table}

\section{The Need for Performance Understanding}
As previously described, much of the existing work has focused on machine
learning performance debugging from the perspective of helping users understand
and improve a model's \emph{prediction
  accuracy}~\cite{modeltracker-amershi15,visinspect-krause16,seq2seqvis-strobelt19,gestalt-patel10,activis-kahng18}.
In this work, we take a complementary approach by focusing on
\emph{computational performance}, specifically for DNN training. In this
section, we provide some background information about DNN training and discuss
some of the common considerations for computational training performance.

\subsection{Background on DNNs and DNN Training}
DNNs, at their heart, are mathematical functions that produce predictions for a
specific task using \emph{learned} parameters---often called
\emph{weights}~\cite{dlbook-goodfellow16}. In practice, DNNs are built by
assembling together a series of \emph{operations}, often called \emph{layers},
each of which may have weights. To make a prediction for an input, a DNN
applies each of its operations in sequence to the input.

A DNN's weights are learned during an iterative process called
\emph{training}~\cite{dlbook-goodfellow16}. In each training iteration, the DNN
processes a \emph{batch} of inputs. Based on the prediction errors it makes,
the DNN's weights are then updated in an attempt to reduce its prediction
error. This process repeats until the DNN reaches an acceptable prediction
accuracy~\cite{dlbook-goodfellow16}.

In practice, DNNs are often trained using hardware accelerators such as
GPUs~\cite{turing-nvidia} or TPUs~\cite{tpu-jouppi17} because they offer
significant performance improvements over CPUs~\cite{mlperf,tpu-jouppi17}. In
this work, we focus on DNN training performance on \emph{GPUs} because they are
\begin{enumerate*}[label=(\roman*)]
  \item more commonly used by deep learning
    developers~\cite{dlbook-goodfellow16}, and
  \item are readily available for purchase and
    rent~\cite{turing-nvidia,gcp-gpus,aws-gpus,azure-gpus}.
\end{enumerate*}

\subsection{Common Performance Considerations}
Although training is a conceptually simple procedure, it can be
computationally-expensive, memory-intensive, and time-consuming in
practice~\cite{mlperf,dawn-coleman17,tbd-zhu18,bert-devlin19,mlenergy-strubell19}.
As a result, there are a number of common considerations for DNN training
performance; we outline and discuss them in more detail below.

\namedparagraph{Training Throughput.}
The key metric for computational performance is the DNN's \emph{training
  throughput}; it indicates how well the underlying hardware is being
utilized~\cite{tbd-zhu18}. Throughput measures how many data samples are
processed per unit of time, and is therefore calculated by dividing the batch
size by the time it takes to run one training iteration.

\namedparagraph{Batch Size.}
The \emph{batch size} is one factor that affects the training throughput. Using
a larger batch size may lead to higher training throughputs on some models, but
usually with diminishing returns~\cite{tbd-zhu18,dawn-coleman17,mlperf}. The
batch size can also affect the \emph{convergence} (i.e. final attainable
prediction accuracy) of the
model~\cite{efficientbp-lecun12,gengap-keskar17,mlperf}. As a result, deep
learning developers need to empirically adjust the batch size to balance
computational performance and convergence. In this work, we focus on guiding
the computational performance aspect of this tuning (i.e. selecting batch sizes
that lead to increased throughput).

\namedparagraph{Memory Usage.}
Memory usage is an important performance consideration because a DNN generally
needs to fit in the memory of the underlying hardware system. A DNN's training
memory footprint consists primarily of memory allocated for the model's
\begin{enumerate*}[label=(\roman*)]
  \item \emph{weights}, and
  \item \emph{activations}.
\end{enumerate*}
Activations are computed in each training iteration and are used to determine
how to update the weights at the end of each
iteration~\cite{dlbook-goodfellow16}.

A natural first step in improving a DNN's computational training performance is
discovering and understanding the DNN's performance bottlenecks. However, as
described previously, this debugging process can still be a difficult endeavor
for the typical deep learning developer.

\section{Performance Understanding: Formative Study}
To better understand how deep learning developers think about and approach
computational performance, we conducted a formative study. We interviewed five
participants who work with DNNs: three deep learning graduate student
researchers (P1, 5 years of experience working with DNNs; P2, 4.5 years; P3,
4.5 years) and two deep learning practitioners in industry (P4, 4 years; P5, 4
years). Four interviews were in-person and one was remote. Each interview took
up to 30 minutes to complete and each participant received a \$25 gift card.

The purpose of our interviews was to learn
\begin{enumerate*}[label=(\roman*)]
  \item about the importance of computational training performance to deep
    learning developers,
  \item how developers currently deal with computational performance, and
  \item whether they face any challenges understanding and or debugging
    performance.
\end{enumerate*}
To elicit this information, we asked each participant open-ended questions
about computational performance. We began by asking them to describe if and how
computational performance influences their models' designs. Then, to learn how
they deal with performance, we asked them about the metrics they use to
evaluate performance and how they obtain these metrics today. Finally, we asked
them to describe any challenges they face when understanding and debugging
performance. We summarize the key takeaways from our interviews below.

\subsection{Performance Constrains Model Design}
P1, P3, P4, and P5 all said that computational performance is important to
them. P2 mentioned that performance is only a concern if it is noticeable,
stating that otherwise the effort needed to find and debug performance issues
would not be worth the time saved---alluding to the difficulty of debugging
performance in the first place.
P1 and P3 noted that performance is important because it \emph{constrains how
  they design their models}. P3 specifically stated that performance affects
which model designs they can experiment with due to the limitations of the
computational resources they have available.

\subsection{Participants Lack Useful Profiling Tools}
The participants mentioned that, for performance understanding and debugging,
they either
\begin{enumerate*}[label=(\roman*)]
  \item do not use any tools at all,
  \item manually instrument their code to measure run time, or
  \item rely on existing profiling tools that are difficult to use or lack
    features such as memory usage breakdowns.
\end{enumerate*}

P1 and P2 said that they do not use \emph{any} tools for understanding
performance and rather just \emph{guess} based on what ``feels slow.'' P2 went
further to say that PyTorch, the software framework they use, hides performance
details through its abstractions---making it difficult to debug performance.
P3 noted that it is impossible to see breakdowns of a model's memory usage and
training iteration run time in existing tools, which they felt to be useful
metrics when investigating performance issues. P5 said that they resort to
manually instrumenting parts of the code that they want to debug.

P4 was the most vocal about the lack of useful and easy to use DNN profiling
tools. Since they use GPUs in their work, they experimented with using
nvprof~\cite{nvprof}: a general-purpose GPU computation performance debugger.
They described their experience with nvprof as a ``disaster'' because it
provided too much low-level GPU information. They had trouble connecting this
low-level information to the high-level PyTorch code that they wrote.  Their
frustration was a result of existing performance tools working at \emph{too
  low} of an abstraction level for typical deep learning developers.

\subsection{Understanding Memory Usage is a Shared Concern}
P1, P2, P3, and P5 said that understanding how memory is used in their models
would be useful. P1 noted that their models tend to use a lot of memory and
that it is difficult to determine what part of the model is responsible for the
high memory usage. Similarly, P2 and P3 said that having a breakdown of their
model's memory usage would be helpful, with P2 noting that a memory breakdown
would help them decide how to shape their model's layers.

P1, P2, and P3 all mentioned that they experiment with different batch sizes
during training to adjust their model's memory usage. P1 went further to say
that they tune the size of their model's layers along with the batch size to
fit within memory constraints, and that having visibility into their model's
memory usage would help with this process.

Overall, these takeaways point to a lack of performance tools that
\begin{enumerate*}[label=(\roman*)]
  \item showcase \emph{useful} performance information (e.g., memory usage
    breakdowns),
  \item are \emph{easy to use}, and
  \item work at the \emph{right abstraction level}.
\end{enumerate*}
As a result, our interviewees resorted to either ignoring performance problems
or expending significant effort to manually instrument their code.

\section{Skyline}\label{sec:skyline}
To help fill this need for new performance profiling tools, we present
\thetool{}: a new interactive in-editor tool for DNN training on GPUs that
supports performance profiling, visualization, and debugging. In this section,
we begin by outlining \thetool{}'s design goals and the special properties of
DNN training computation that enable \thetool{} to achieve these goals. Then,
we describe \thetool{}'s features in detail.

\begin{figure*}
  \centering
  \subfigure{%
    \begin{overpic}[width=0.95\textwidth]{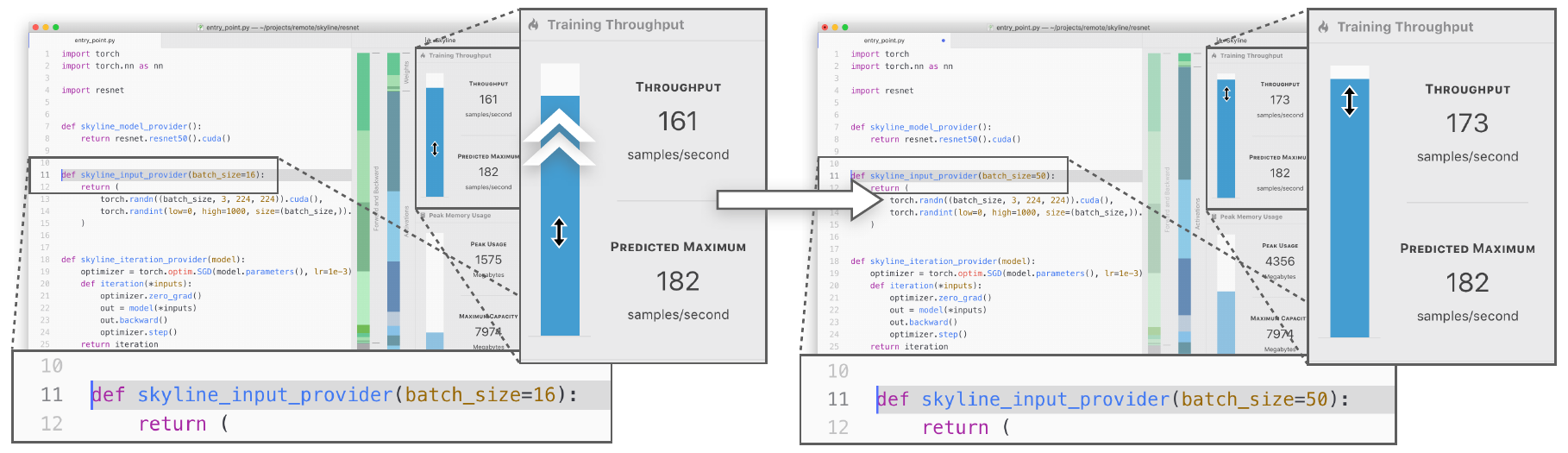}
      \put(30,19){\captionlabel{A}}
      \put(37,3.1){\captionlabel{B}}
      \put(96.1,21){\captionlabel{D}}
      \put(87,3.1){\captionlabel{C}}
      \put(71.5,10){\captionlabel{E}}
    \end{overpic}
    \label{fig:draggable-drag-up}
    \addtocounter{subfigure}{1}
    \label{fig:draggable-orig-batch}
    \addtocounter{subfigure}{1}
    \label{fig:draggable-new-batch}
    \addtocounter{subfigure}{1}
    \label{fig:draggable-new-thpt}
    \addtocounter{subfigure}{1}
    \label{fig:draggable-new-other}
  }
  \caption{An example of \thetool{}'s code-linked manipulatable key performance
    visualizations.
    \subref{fig:draggable-drag-up}~Users can drag the bar charts, e.g.,
      upwards.
    \subref{fig:draggable-orig-batch}~The initial batch size in the code.
    \subref{fig:draggable-new-batch}~Dragging the throughput bar leads to a new
    batch size that is predicted to result in the dragged throughput.
    \subref{fig:draggable-new-thpt}~The new throughput, after dragging.
    \subref{fig:draggable-new-other}~Other visualizations (memory breakdown and
      usage) update when the throughput bar is dragged, and vice versa.
  }
  \label{fig:draggable}
  \vspace{-1em}
\end{figure*}

\subsection{Skyline Design Goals}
\thetool{}'s design is guided by the following goals, which encompass the
insights from our formative study.

\subsubsection{Goal 1 (G1): Appropriate Level of Abstraction}
\thetool{} needs to present performance information at the right level of
abstraction for everyday deep learning developers. These users develop their
models in software frameworks (e.g., PyTorch) and are therefore most familiar
with the abstractions used in these frameworks.

\subsubsection{Goal 2 (G2): Code-Connected Insights}
Performance profiling tools should help deep learning developers uncover
performance issues in their code. This means that \thetool{} should make it
easy for users to identify the relevant lines of code associated with any
performance information it displays (e.g., visualizations of operation run
times).

\subsubsection{Goal 3 (G3): Relevant Metrics}
\thetool{} needs to focus on surfacing relevant metrics that deep learning
developers care about for DNN training. These metrics include training
throughput, iteration run time, and overall memory usage. This is because these
metrics are directly affected by
\begin{enumerate*}[label=(\roman*)]
  \item the code that the user writes, and
  \item parameters that the user sets (e.g., the batch size).
\end{enumerate*}

\subsection{What Makes DNN Training Computation Special?}
\vspace{0.05em}
\subsubsection{Property 1 (P1): Fast and Repetitive Iterations}
Although DNNs can take hours or days to train to an acceptable
accuracy~\cite{mlperf,tbd-zhu18,dawn-coleman17}, the training process consists
of repetitions of a \emph{single} training
iteration~\cite{dlbook-goodfellow16}, which runs on the order of hundreds of
\emph{milliseconds}~\cite{tbd-zhu18}.
Additionally, for many DNNs, the computation in a training
iteration depends only on the \emph{size} of the inputs, which is fixed
throughout training\footnote{Language-based models (e.g.,
  GNMT~\cite{gnmt-wu16}) take variable-sized inputs because they operate on
  sentences. For these kinds of models, \thetool{} can use the longest sentence
  in the dataset as input during profiling to provide a lower-bound on the
  measured performance.} (i.e. generally, the same ``amount'' of computation
runs regardless of the values of the inputs).
As a result, the computational performance of an entire training process can be
characterized by the performance of just a few training iterations.

\subsubsection{Property 2 (P2): Predictability for Batch Sizes}
The performance (throughput, memory footprint) of a training iteration with
respect to its batch size is predictable, as we describe and show in more
detail in the \hyperref[sec:impl]{System Implementation} section. This property
helps enable the directly manipulatable visualizations in \thetool{}.

\subsubsection{Property 3 (P3): Structured Code}
DNNs are built using software frameworks (e.g., PyTorch~\cite{pytorch-paszke19}
or TensorFlow~\cite{tensorflow-abadi16}), which offer a set of abstractions
that \emph{constrain} how a model can be implemented (e.g., all models in
PyTorch are instances of a specific class). This means DNN code is written in a
structured way, which helps \thetool{} connect performance visualizations to
specific lines of code.

\subsection{Overview}
\thetool{} runs as a plugin in Atom~\cite{atom} (a popular open-source editor)
and supports DNNs implemented using PyTorch~\cite{pytorch-paszke19}. It runs
within the editor, in a sidebar to the right of an opened file
(Figure~\ref{fig:overview}). Although we implemented \thetool{} to work with
Atom and PyTorch, the ideas behind it are \emph{not} fundamentally limited to
this editor and framework combination. We discuss extensibility in more detail
in our \hyperref[sec:discussion]{Discussion} section.

To use \thetool{}, users first write an entry point file that contains special
\emph{provider functions} that tell \thetool{} how to run their model. We
describe these functions in more detail in the
\hyperref[sec:skyline-providers]{Provider Functions} section. Then, users start
\thetool{} by navigating to the directory containing their entry point file and
running a \thetool{} command in their terminal.

One of \thetool{}'s key features is that it provides \emph{live} performance
feedback. When users make and save changes to their model in the editor,
\thetool{} re-profiles their code in the background and updates the
visualizations with the latest performance data. \thetool{} can offer live
feedback because getting performance data about a model only requires profiling
a few short training iterations (\emph{P1}).

\subsection{Interactive Key Performance Metrics}
\thetool{} uses bar charts to visualize a model's training throughput and peak
memory usage (\emph{G1, G3}). Respectively, these charts show the current
throughput and peak memory usage relative to the maximum achievable throughput
(predicted) and memory capacity on the user's GPU
(Figure~\ref{fig:draggable}\subref{fig:draggable-drag-up}).

One of \thetool{}'s novel contributions is the ability for the user to
\emph{directly manipulate} these bar charts (Figures~\ref{fig:overview-predict}
and \ref{fig:draggable}). As the user drags the bars up or down
(Figure~\ref{fig:draggable}\subref{fig:draggable-drag-up}), \thetool{} makes
\emph{predictions} about the batch sizes that can be used during training to
achieve these manipulated metrics. While the dragging occurs, \thetool{} also
\begin{enumerate*}[label=(\roman*)]
  \item \emph{mutates} the user's code to actually set these predicted batch
    sizes (Figure~\ref{fig:draggable}\subref{fig:draggable-new-batch}), and
  \item updates the other visualizations to account for the changed batch size
    (Figure~\ref{fig:draggable}\subref{fig:draggable-new-other}).
\end{enumerate*}
For example, if the user drags the throughput bar upwards, \thetool{} will
predict that a larger batch size needs to be used. \thetool{} would then update
the memory footprint breakdown (described in the following section) to show
that activations will comprise a larger proportion of the memory footprint.
This feature is made possible by the predictability of DNN training
computational performance and the structured nature of the code (\emph{P2,
  P3}).

In addition to computational performance, the batch size used during training
can also affect a model's final prediction accuracy~\cite{gengap-keskar17}.
While the effects of the batch size on a model's quality are generally
non-trivial to predict~\cite{dpeffects-shallue19} (see our
\hyperref[sec:discussion]{Discussion} section), small changes to the batch size
empirically do not seem to have a significant effect on a model's final
prediction accuracy. To err on the side of caution, \thetool{} does not make
model accuracy predictions and it does not automatically recommend batch sizes
to use based on its performance predictions. \thetool{} instead offers these
manipulatable visualizations to allow users to explore the computational
performance trade-offs of different batch sizes. We envision this feature as a
way for developers to become more informed about the performance implications
of their batch size selections.

\subsection{Interactive Breakdowns}
\thetool{} also displays detailed breakdowns of the \emph{training iteration
  run time} and \emph{memory footprint}, visualizing them using stacked bar
charts (Figure~\ref{fig:breakdown-details}). The bars in each bar chart are
sorted in descending order based on their size.
These breakdowns indicate how each operation contributes to the iteration run
time (Figure~\ref{fig:breakdown-details}\subref{fig:breakdown-details-runtime})
and memory footprint. The memory footprint breakdown is further split between
the memory used for the activations and weights
(Figures~\ref{fig:breakdown-details}\subref{fig:breakdown-details-weights}
and~\ref{fig:breakdown-details}\subref{fig:breakdown-details-activations}).

\subsubsection{Interactivity}
\begin{figure}
  \centering
  \subfigure{%
    \begin{overpic}[width=\columnwidth]{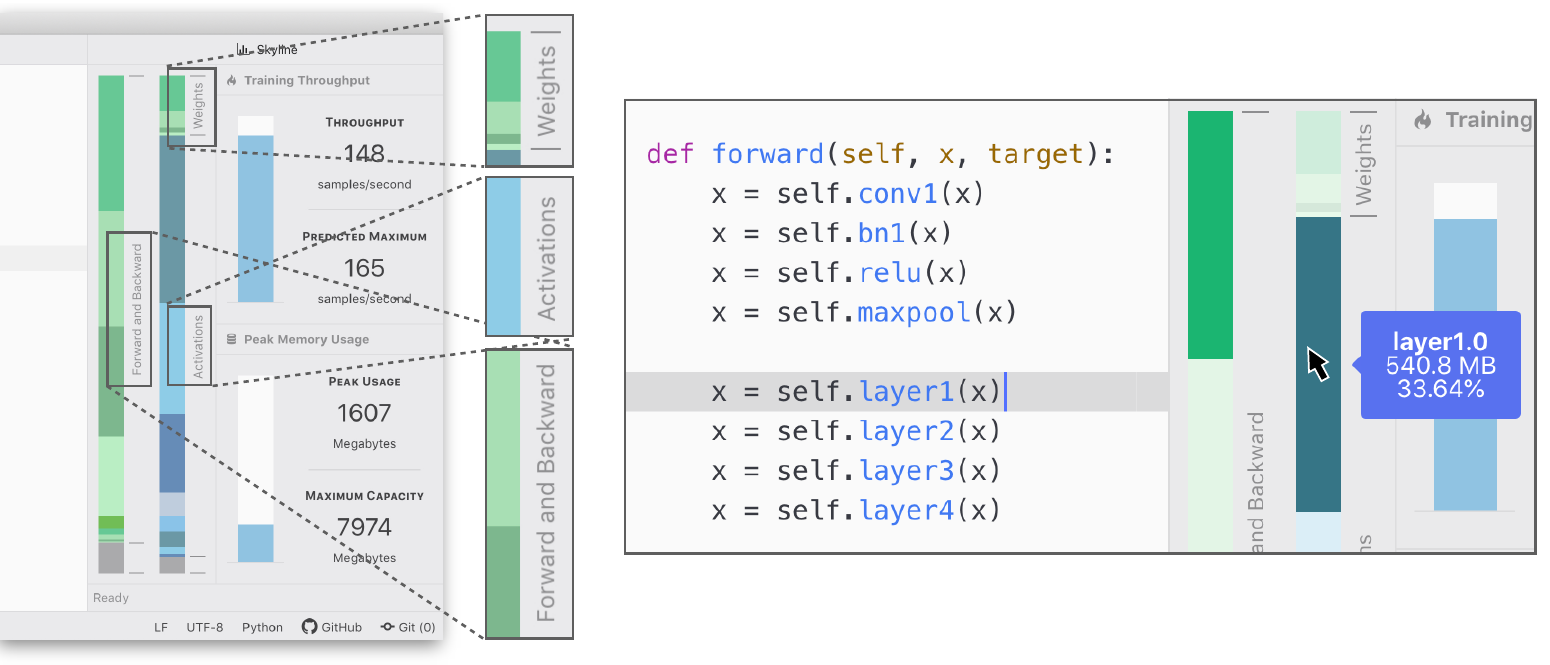}
      \put(24,35.5){\captionlabel{A}}
      \put(24,25.5){\captionlabel{B}}
      \put(24,10.5){\captionlabel{C}}
      \put(78,25){\captionlabel{D}}
      \put(67,16.5){\captionlabel{E}}
    \end{overpic}
    \label{fig:breakdown-details-weights}
    \addtocounter{subfigure}{1}
    \label{fig:breakdown-details-activations}
    \addtocounter{subfigure}{1}
    \label{fig:breakdown-details-runtime}
    \addtocounter{subfigure}{1}
    \label{fig:breakdown-details-hover}
    \addtocounter{subfigure}{1}
    \label{fig:breakdown-details-highlight}
  }
  \vspace{-1em}
  \caption{An overview of \thetool{}'s interactive breakdowns.
    \subref{fig:breakdown-details-weights} Memory used by weights.
    \subref{fig:breakdown-details-activations} Memory used by activations.
    \subref{fig:breakdown-details-runtime} Operation run times.
    \subref{fig:breakdown-details-hover} Breakdowns bars are linked; hovering
      over a memory bar will highlight the corresponding run time bar and vice
      versa.
    \subref{fig:breakdown-details-highlight} Relevant line(s) of code are
      highlighted when hovering over visualizations.
  }
  \label{fig:breakdown-details}
  \vspace{-1.2em}
\end{figure}

When a user hovers over one of these bars, \thetool{}
\begin{enumerate*}[label=(\roman*)]
  \item highlights the relevant line of code associated with that bar
    (Figure~\ref{fig:breakdown-details}\subref{fig:breakdown-details-highlight})
    (\emph{G2, P3}), and
  \item reveals details about the bar: its run time or memory usage as a
    concrete number and as a percentage of the total iteration run time or
    memory usage (Figure~\ref{fig:overview-hover}).
\end{enumerate*}
If the user clicks the bar, \thetool{} will place their cursor on the bar's
associated line of code and will also open the file containing that line of
code in the editor.

\subsubsection{Navigating the Breakdown Hierarchy}
A DNN may contain hundreds of operations, which can be overwhelming when
visualized all at once. To manage this complexity, \thetool{} organizes a DNN's
operations into a \emph{hierarchy of modules} where the top-level module
represents the entire DNN. Each module is a ``container'' that can hold
operations as well as other modules (analogous to a directory in a file
system). \thetool{} builds this hierarchy by leveraging the observation that
deep learning developers actually implement DNNs using modules; they are an
abstraction used in PyTorch (\emph{P3}). \thetool{} uses stack traces, recorded
during profiling, to extract the hierarchy from the user's code; we discuss
this process in more detail in the \hyperref[sec:impl]{System Implementation}
section.

\thetool{}'s breakdowns show the contents of one module at a time, starting
with the top-level module representing the entire DNN (\emph{G1}). Because of
this design, some bars in the breakdowns will represent a module and will
visualize the aggregate of all the operations inside it (e.g., the sum of all
the run times of the operations in that module). When the user \emph{double
  clicks} these bars, \thetool{} will ``expand'' the bar and visualize the
operations \emph{inside} that module. \thetool{} has buttons in its user
interface that allow users to
\begin{enumerate*}[label=(\roman*)]
  \item go back one level up, and
  \item return to the top of the hierarchy.
\end{enumerate*}

These hierarchical breakdowns allow the user to \emph{navigate} the performance
profile of their model by drilling down to the specific parts of the model that
they are interested in. For example, if a user wanted to investigate run time
bottlenecks, they could double click the largest run time breakdown bar to
drill down to the operations responsible for the large run time.

\subsection{Inline Performance Markers}
\begin{figure}
  \centering
  \subfigure{%
    \begin{overpic}[width=\columnwidth]{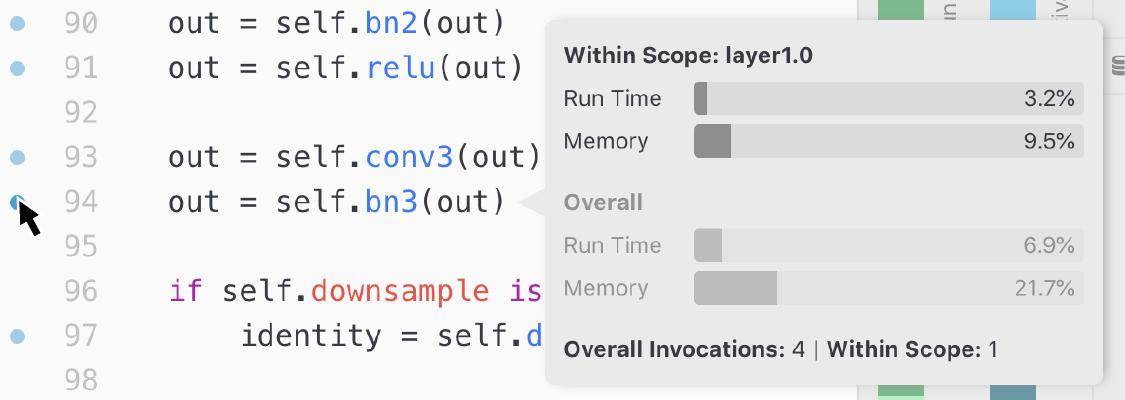}
      \put(41,24){\captionlabel{A}}
      \put(41,10){\captionlabel{B}}
    \end{overpic}
    \label{fig:inline-code-scope}
    \addtocounter{subfigure}{1}
    \label{fig:inline-code-overall}
  }
  \caption{Inline markers stay in-sync with the current breakdown module being
    explored.
    \subref{fig:inline-code-scope} Run time and memory usage specific to module
    ``\texttt{layer1}.''
    \subref{fig:inline-code-overall} Performance details for all invocations of
    this line of code.}
  \label{fig:inline-code}
  \vspace{-1.2em}
\end{figure}

\thetool{} places inline markers in the editor gutter to indicate the lines of
code (which correspond to operations in the model) for which performance
information is available (Figure~\ref{fig:overview-inline}). When hovering over
one of these markers, \thetool{} displays detailed information about that line
of code: the amount it contributes to the iteration run time and memory
footprint. These visualizations are also linked to the state of the breakdowns.
For example, if the user double clicks a breakdown bar to view the contents of
a module, the inline markers will show performance details about that specific
module as well (Figure~\ref{fig:inline-code}). This linking is useful because
modules are defined once in the code but may be used multiple times in a DNN.

Having both inline markers and breakdowns allows for two different approaches
to performance investigation. If the user does not know what their model's
bottlenecks are, they can use the breakdowns to explore and drill down to
specific parts of their model. But, if they already have concerns about
specific lines of code, they can navigate directly to that code and inspect the
inline markers for detailed performance information.

\subsection{Provider Functions}\label{sec:skyline-providers}
\thetool{} gathers performance data about the user's model by actually
\emph{running} the model on a set of chosen inputs. For this to work, users
need to implement three \thetool{} \emph{provider functions}. In a nutshell,
these providers are
\begin{itemize}[noitemsep]
  \item \textit{Model Provider:} Returns an instance of the user's model.
  \item \textit{Input Provider:} Returns inputs for the user's model.
  \item \textit{Iteration Provider:} Returns a function that, when invoked,
    runs a single training iteration.
\end{itemize}
\thetool{} uses these providers, which are implemented in Python, to run the
user's model during profiling. We designed the providers to be easy to
implement for a user who is familiar with the model's code. We also evaluated
the difficulty of implementing these providers in our user study (see the
\hyperref[sec:user-eval]{User Evaluation} section).

\section{System Implementation}\label{sec:impl}
\subsection{Software Architecture}
\thetool{} consists of two components:
\begin{enumerate*}[label=(\roman*)]
  \item a \emph{plugin} that runs in Atom that displays all of \thetool{}'s
    visual and interactive features, and
  \item a \emph{daemon process} that performs the actual performance profiling.
\end{enumerate*}
The two components communicate over a network socket using a
\thetool{}-specific protocol. This architecture enables users to do their
development on one machine (e.g., a laptop running Atom with the \thetool{}
plugin) while profiling their code on a different machine (e.g., a virtual
machine in the cloud).\footnote{This remote setup is necessary when users need
  to train models on remote resources (e.g., a shared GPU compute cluster).
  \thetool{} also supports developing and profiling on the same machine.}

\subsection{Gathering Performance Data}
Each time the user modifies and saves their code, \thetool{} runs their model
in the background to profile it and gather performance data (throughput, memory
usage, operation run times). \thetool{} ``monkey patches'' PyTorch operations
with wrappers that allow it to intercept and keep track of all the operations
that run in one training iteration, as they are executed.

\namedparagraph{Profiling Time.}
The time it takes for this profiling process to complete depends on the DNN,
batch size, and underlying GPU. For example, with the experimental setup
described in the \hyperref[sec:evaluate-prediction-accuracy]{Evaluation of
  Prediction Accuracy} section, profiling a feed-forward DNN with three hidden
layers would take less than one second whereas profiling the Transformer would
take up to 29 seconds. \thetool{}'s profiling runs in the background and does
\emph{not} interfere with the user's ability to write code in their editor. If
desired, users can also disable this ``re-profile on save'' feature and instead
manually trigger profiling by clicking a button on \thetool{}'s user interface.

\namedparagraph{Measuring Throughput.}
To measure the training throughput, \thetool{} measures the time it takes to
run three training iterations, to account for any variance in the run times.
\thetool{} then computes the throughput by dividing three times the input batch
size by the measured run time. \thetool{} uses the user-written provider
functions to extract the user's desired batch size and to run the user's
training iteration code.

\namedparagraph{Measuring Operation Run Times.}
When \thetool{} intercepts an operation, it runs the operation independently to
measure its run time. This approach allows \thetool{} to
\begin{enumerate*}[label=(\roman*)]
  \item decompose the overall iteration run time into individual operations,
    and
  \item amortize any profiling overhead by running each operation multiple
    times.
\end{enumerate*}
These measured run times are used by the run time breakdown stacked bar chart
visualization.

\namedparagraph{Measuring Memory.}
\thetool{} explicitly tracks memory allocated for the model's
\begin{enumerate*}[label=(\roman*)]
  \item \emph{weights}, and
  \item computed \emph{activations}.
\end{enumerate*}
\thetool{} tracks the memory allocated for the weights by intercepting all
weight creations when a model is instantiated. The memory used for activations
is allocated as each DNN operation runs during a training iteration. To track
this memory, \thetool{} records the additional memory consumed by each
operation after it runs. These measurements are used by the memory breakdown
stacked bar chart visualization.

\namedparagraph{Untracked Run Time and Memory.}
\thetool{} only explicitly tracks run times and memory allocations that are
attributed to weights or operations in a DNN. For completeness, the remaining
run time and memory usage are displayed as ``untracked'' bars in the run time
and memory breakdown. These untracked bars represent auxiliary tasks that run
during a training iteration (e.g., weight updates, memory allocated by the
underlying framework for bookkeeping). They typically do not comprise a
significant proportion of the overall iteration run time and memory usage (less
than 20\% in our use).

\namedparagraph{Code References.}
When an operation is executed, the state of the call stack contains all the
relevant lines of code leading up to running the operation. Thus, when an
operation is intercepted, \thetool{} also records the file and line number for
each frame in the call stack. This allows \thetool{} to connect the operation's
performance data to the relevant line(s) of code.

\namedparagraph{Hierarchical Breakdowns.}
\thetool{} uses the \emph{stack trace} associated with each performance
measurement to assemble the measurements into a hierarchy that mirrors the
modules used in the code. This approach works because
\begin{enumerate*}[label=(\roman*)]
  \item all the operations in the model share a common stack frame, which is
    where the model is invoked in the code; and
  \item operations in different modules are defined in different functions,
    meaning any two operations in different modules will always appear in
    different stack frames.
\end{enumerate*}
\thetool{} assembles all the stack traces into a tree where nodes represent
operations or modules. If a node has children, they represent the operations
and modules contained inside that node. \thetool{}'s breakdowns visualize the
children of one node at a time, starting with the root node (i.e. the
operations and modules in the top-level module). Leaf nodes always represent
the operations in the model.

\begin{figure*}
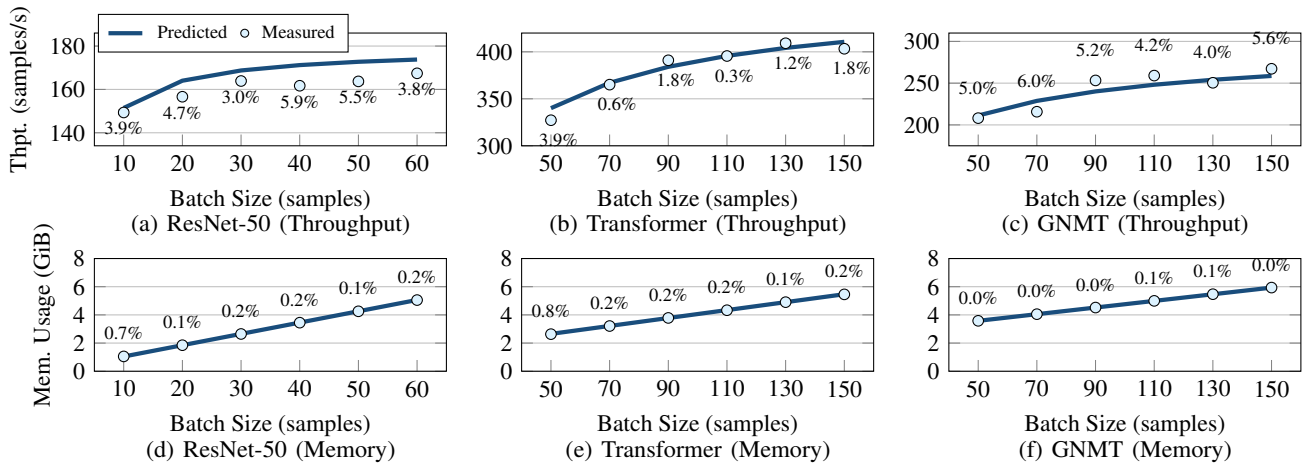

  \centering
  \subfigure{%
    \begin{tikzpicture}[baseline]
    \begin{groupplot}[
      group style = {
        group name=pred,
        group size=3 by 2,
        vertical sep=1.5cm,
      }
    ]
    \input{charts/predictions-resnet-thpt}
    \input{charts/predictions-transformer-thpt}
    \input{charts/predictions-gnmt-thpt}
    \input{charts/predictions-resnet-mem}
    \input{charts/predictions-transformer-mem}
    \input{charts/predictions-gnmt-mem}
    \end{groupplot}
    \node[text width=6cm,align=center,anchor=north] at
      ([yshift=-8mm]pred c1r1.south)
      {\footnotesize \subref{fig:predictions-resnet-thpt}~ResNet-50 (Throughput)\label{fig:predictions-resnet-thpt}};
    \addtocounter{subfigure}{1}
    \node[text width=6cm,align=center,anchor=north] at
      ([yshift=-8mm]pred c2r1.south)
      {\footnotesize \subref{fig:predictions-transformer-thpt}~Transformer (Throughput)\label{fig:predictions-transformer-thpt}};
    \addtocounter{subfigure}{1}
    \node[text width=6cm,align=center,anchor=north] at
      ([yshift=-8mm]pred c3r1.south)
      {\footnotesize \subref{fig:predictions-gnmt-thpt}~GNMT (Throughput)\label{fig:predictions-gnmt-thpt}};
    \addtocounter{subfigure}{1}
    \node[text width=6cm,align=center,anchor=north] at
      ([yshift=-8mm]pred c1r2.south)
      {\footnotesize \subref{fig:predictions-resnet-mem}~ResNet-50 (Memory)\label{fig:predictions-resnet-mem}};
    \addtocounter{subfigure}{1}
    \node[text width=6cm,align=center,anchor=north] at
      ([yshift=-8mm]pred c2r2.south)
      {\footnotesize \subref{fig:predictions-transformer-mem}~Transformer (Memory)\label{fig:predictions-transformer-mem}};
    \addtocounter{subfigure}{1}
    \node[text width=6cm,align=center,anchor=north] at
      ([yshift=-8mm]pred c3r2.south)
      {\footnotesize \subref{fig:predictions-gnmt-mem}~GNMT (Memory)\label{fig:predictions-gnmt-mem}};
    \end{tikzpicture}
  }
  \vspace{-0.8em}
  \caption{An evaluation of \thetool{}'s predictive models for training
    throughput and memory usage. Each graph depicts \thetool{}'s throughput or
    memory usage prediction with a solid line. The prediction error for
    specific batch sizes is shown as a percentage next to each measured
    point.}\label{fig:predictions}
  \vspace{-1em}
\end{figure*}

\subsection{Making Performance Predictions}
\thetool{}'s interactive features link a DNN's throughput and memory usage to
its batch size using predictive models. These predictive models help users tune
their DNN's batch size when they drag the throughput or memory usage
visualizations (Figure~\ref{fig:overview-predict}). To stay up-to-date,
\thetool{} builds new predictive models each time the user modifies and saves
their code.

\subsubsection{Throughput}
Let $x$ represent the batch size and let $a$ and $b$ be constants. \thetool{}
models the training throughput, $T(x)$, and iteration run time, $R(x)$, using
\begin{align}
  T(x) &= \frac{x}{R(x)} \\
  R(x) &= ax + b
\end{align}
The intuition behind this model is that, as the underlying hardware is
saturated, increasing the number of inputs to process will result in a linear
increase in the amount of time it takes process them ($R(x)$). In the limit
(i.e. as $x$ approaches infinity), the throughput approaches $1/a$, which is
what \thetool{} uses to predict the maximum attainable throughput.

\thetool{} selects $a$ and $b$ empirically by measuring $R(x)$ for three
different values of $x$ (i.e. three different batch sizes) and then applying
least squares linear regression~\cite{prml-bishop06}. \thetool{} uses the
batch size set by the user in the input provider function as the first batch
size, and then selects the other two batch sizes by adding a constant to the
previously sampled batch size. If a selected batch size results in running out
of memory, \thetool{} will adjust and select smaller batch sizes
instead.\footnote{In the rare case where a predictive model cannot be built
  because the user's DNN is too large to support three different batch sizes,
  the ability to drag the throughput and memory bar charts is disabled.}

\subsubsection{Memory Usage}
Let $x$ represent the batch size and let $c$ and $d$ be constants. \thetool{}
models the overall memory usage, $M(x)$, using
\begin{align}
  M(x) = cx + d
\end{align}
The intuition behind this model is that the memory used by the computed
activations is directly proportional to the batch size because there exist
activations for each sample in the batch. Just like for the throughput model,
\thetool{} selects $c$ and $d$ empirically by measuring $M(x)$ for three
different values of $x$ and then applying least squares linear regression.
\thetool{} applies the same strategy for selecting batch sizes to sample as for
the throughput model.

\subsubsection{Using the Predictive Models}
\thetool{} uses these models when the user manipulates the key performance
metric bar charts (throughput and memory usage). For example, when the user
drags the memory usage bar chart, \thetool{} converts the distance dragged into
a new memory usage value. \thetool{} then translates this memory usage into a
batch size and throughput prediction by using the memory predictive model
followed by the throughput predictive model. The throughput prediction is then
used to update the throughput bar chart. \thetool{} takes a similar approach
when the user drags the throughput bar chart.

To mutate the code, \thetool{} uses Python's built-in abstract syntax tree
parser to identify the line of code that contains the input provider function's
signature. The user defines the batch size to use by setting a keyword argument
in the input provider's signature. \thetool{} mutates this line of code by
updating the signature to reflect the new predicted batch size.

\subsection{Evaluation of Prediction
  Accuracy}\label{sec:evaluate-prediction-accuracy}
We empirically evaluate the effectiveness of these predictive models by
measuring their \emph{prediction errors} (lower is better) for training
throughput and memory usage across three different DNNs and six different batch
sizes.

\namedparagraph{Experimental Setup.}
We run our experiments on a machine equipped with
\begin{enumerate*}[label=(\roman*)]
  \item an AMD Ryzen TR 1950X 3.4 GHz 16-core CPU~\cite{amd-1950x},
  \item 16 GB of DDR4 main memory~\cite{ddr4}, and
  \item an NVIDIA GeForce RTX 2070 GPU~\cite{2070} with 8 GB of
    GDDR6 memory~\cite{gddr6}.
\end{enumerate*}
We use PyTorch 1.3.1~\cite{pytorch-paszke19} and CUDA 10.1~\cite{cuda},
running on Ubuntu 18.04~\cite{ubuntu}.

\namedparagraph{Methodology.}
We use \thetool{} to build predictive models for training throughput and memory
usage for ResNet-50~\cite{resnet-he16}, the
Transformer~\cite{attention-vaswani17}, and GNMT~\cite{gnmt-wu16}, starting
from three different batch sizes for each model (8, 16, and 32 for ResNet-50;
32, 64, and 80 for the Transformer and GNMT). These models are used in
well-known DNN benchmarking suites including MLPerf~\cite{mlperf} and
TBD~\cite{tbd-zhu18}. The ResNet-50 model is configured for the ImageNet
dataset~\cite{imagenet} and the Transformer and GNMT are configured for the
WMT'16 (EN-DE) dataset~\cite{wmt16}, with fixed sentences of length 25. We use
synthetic training data for all three models because we are evaluating the
\emph{computational} performance of the models, which we verified to not
depend on the values of the input data.

We then use the predictive models to make training throughput and memory usage
predictions for six other batch sizes (shown in Figure~\ref{fig:predictions}).
We compare the predictions against the measured throughput and memory usage.
When measuring the throughput and memory usage, we repeat each measurement five
times and use the median measurement. To amortize any measurement overhead when
measuring the throughput, we record the time it takes to run three training
iterations and divide three times the batch size by this run time.

\namedparagraph{Results.}
Figure~\ref{fig:predictions} shows a summary of \thetool{}'s prediction
accuracy results. Since we create three predictive models for each DNN
(starting from different batch sizes) and make predictions using them for all
six different batch sizes, we average the prediction values and prediction
errors across all three predictive models when plotting them in
Figure~\ref{fig:predictions}. Next to each measured point, we show the average
prediction error for that batch size. The standard deviations of all our
throughput and memory usage measurements are under \ThptMaxStdDev{} and
\MemMaxStdDev{} respectively.

From this figure we can draw two major conclusions. First, \thetool{} makes
\emph{accurate} throughput and memory usage predictions. The average error for
throughput and memory usage among the three DNNs we evaluate is \ThptAvgError{}
and \MemAvgError{} respectively (maximum \ThptMaxError{} and \MemMaxError{}).
Second, \thetool{}'s predictive models work across a \emph{diverse} set of DNN
architectures and tasks. ResNet-50 is a convolutional neural network used for
image classification, the Transformer is an attention-based DNN used for
translation, and GNMT is a recurrent neural network also used for translation.

\section{User Evaluation}\label{sec:user-eval}
To evaluate \thetool{}, we conducted a qualitative user study examining
\thetool{}'s diagnostic capabilities. \thetool{} is not designed for novice
deep learning developers, and so we required the participants to be experienced
with PyTorch (the deep learning framework that \thetool{} supports). This
requirement made it difficult to recruit a large number of qualified
participants. We recruited seven participants (six male), all deep learning
graduate student researchers that train DNNs in their day-to-day work. The
participants were between the ages of 23 and 30 (median 24), and had one to
three years of experience with PyTorch (median 2.5 years). Six of the
participants were affiliated with one institution, and the seventh was
affiliated with a different institution. Two participants had also taken part
in our formative study. We conducted our study sessions remotely using screen
sharing software and compensated participants with a \$25 gift card.

\subsection{Study Design}
We split our study into three parts:
\begin{enumerate*}[label=(\roman*)]
  \item a walkthrough of \thetool{}'s features,
  \item a series of timed tasks, and
  \item a qualitative questionnaire about the usefulness and ease of use of
    \thetool{}.
\end{enumerate*}
Each study session took up to one hour to complete.

\namedparagraph{Walkthrough.}
During the walkthrough, we introduced \thetool{} and demonstrated each of its
features. We gave participants a chance to try each feature for themselves and
to ask questions during this process. The full walkthrough took up to 20
minutes to complete and we conducted it with a ResNeXt-50~\cite{resnext-xie17}
model loaded in \thetool{}.

\namedparagraph{Timed Tasks.}
After the walkthrough, we asked the participants to complete five timed tasks:
\begin{enumerate}[noitemsep]
  \item Find the top-level module or operation that contributes the most to the
    iteration run time. If it is a module, repeat recursively.
  \item Repeat the same as above, but for the module or operation that
    contributes the most to the memory footprint.
  \item Determine whether the weights or activations contribute the most to the
    model's memory footprint.
  \item Find the batch size at which the weights and activations take up an
    equal proportion of the overall memory footprint.
  \item Determine whether increasing the batch size leads to increased training
    throughput but with diminishing returns. If it occurs, after which batch
    size?
\end{enumerate}
We asked the participants to complete these five tasks for three different
DNNs:
\begin{enumerate*}[label=(\roman*)]
  \item GNMT~\cite{gnmt-wu16},
  \item the Transformer~\cite{attention-vaswani17}, and
  \item ResNet-50~\cite{resnet-he16}.
\end{enumerate*}
We designed these tasks with two goals in mind:
\begin{enumerate*}[label=(\roman*)]
  \item to mimic common performance debugging tasks to evaluate the ease of
    using \thetool{} to accomplish them, and
  \item to evaluate whether the participants could independently determine
    which \thetool{} features to use for each task.
\end{enumerate*}

For the first two models, we provided implementations of the provider
functions. To evaluate the difficulty of writing the providers, we asked the
\emph{participant} to implement the providers for the third model. We provided
them with
\begin{enumerate*}[label=(\roman*)]
  \item a copy of the relevant documentation, and
  \item an example of the providers written for a different model.
\end{enumerate*}
This is the same amount of information made available to all new \thetool{}
users.

We did not compare the time it took to complete these tasks in \thetool{}
against other existing profiling tools for DNN training. This is because
\thetool{} is unique in its design; \thetool{} is an interactive in-editor tool
whereas existing tools are either command-line-based~\cite{dlprof,pyprof}, not
domain-specific~\cite{nvprof}, or are not interactive and do not offer
comparable features (e.g., performance predictions, memory
profiling)~\cite{dlprof,pyprof,tensorflowprofiler,nvprof}.  As a result, it
would be difficult or impossible for participants to accomplish these tasks
using any existing profiling tools within the time allocated for the study.

\namedparagraph{Questionnaire.}
After completing the timed tasks, the participants filled out a 10-minute
qualitative questionnaire. We asked participants to rate whether they found
each of \thetool{}'s features to be useful and easy to use. We also asked
additional questions about \thetool{} as a whole, which we discuss in the
\hyperref[sec:eval-results]{Results} section. We used a five-point Likert scale
for each question (1 $\Rightarrow$ ``Strongly Disagree'', 5 $\Rightarrow$
``Strongly Agree'') where ``Strongly Agree'' was always the most positive
option. For example, participants would have to select how strongly they agreed
with the statement that \textit{``Skyline's memory breakdown is a useful
  feature.''}

\subsection{Results}\label{sec:eval-results}
\begin{table}
  \centering
  \footnotesize
  \begin{tabularx}{\columnwidth}{llXXXXX}
    \toprule
    & & \multicolumn{5}{c}{\textbf{Mean Completion Time (seconds)}} \\
    \cmidrule{3-7}
    \textbf{Model} & \textbf{Weights} & T1 & T2 & T3 & T4 & T5 \\
    \midrule
    GNMT & 161 M & 52.0 & 36.4 & 26.0 & 43.6 & 63.7 \\
    Transformer & 94 M & 47.4 & 49.6 & 6.4 & 21.7 & 32.0 \\
    ResNet-50 & 26 M & 28.0 & 29.1 & 5.0 & 22.3 & 23.3 \\
    \bottomrule
  \end{tabularx}
  \caption{Summary of mean completion times for tasks 1 to 5 (T1 -- T5). For a
    sense of their size, we include the number of weights in each model.}
  \label{tbl:task-times}
  \vspace{-1.5em}
\end{table}

Overall, the results of our exploratory study were positive and encouraging.
When qualitatively asked about \thetool{} as a whole, participants unanimously
either agreed or strongly agreed that \thetool{} would be useful (mean 4.7/5;
median 5/5) and that it was easy to use (4.7/5; 5/5).

\subsubsection{Timed Tasks}
Table~\ref{tbl:task-times} shows a summary of the mean task completion time, in
seconds, among all participants for each model and task. These results show
that, on average, each task could be completed within roughly one minute. Our
longest recorded time was 2 minutes and 34 seconds for task 4 on GNMT.

During the study, five out of the seven participants completed all the tasks
independently. The two other participants knew ``what'' they needed to do, but
needed reminders about how to use \thetool{}'s features. For task 1, one
participant knew that they needed to ``drill down'' to different parts of the
model using the breakdowns, but forgot that they needed to double click on a
breakdown bar to do so. For task 4, both participants knew that they needed to
drag something to have \thetool{} make predictions, but needed a reminder to
drag the memory usage and throughput bar charts. We feel that these results are
encouraging given that, with just a brief 20-minute walkthrough,
\begin{enumerate*}[label=(\roman*)]
  \item all the participants had some intuition about how to use \thetool{}
    to complete the debugging tasks, and
  \item most of the participants (5/7) completed the tasks independently.
\end{enumerate*}

\subsubsection{Writing the Providers}
On average, implementing the \thetool{} provider functions took the
participants 7 minutes and 41 seconds (median 7 minutes and 52 seconds). The
longest time was 12 minutes and 48 seconds; the shortest time was 3 minutes and
47 seconds. Of the seven participants, four implemented the providers
independently whereas three needed help with minor syntax or PyTorch API
issues (missing commas or keyword arguments, being unsure about the signature
of a PyTorch function).
When asked qualitatively about whether they agreed that writing the provider
functions was easy, two participants strongly agreed, two agreed, and three
were neutral.
We believe that these results are still encouraging given that this was the
first time each of the participants tried to implement the \thetool{}
providers. Participant~1 commented that \textit{``[the providers] were very
  little setup for the benefits of Skyline.''} Similarly, participant 3
mentioned that implementing the providers was easier for them than the
equivalent setup they had to perform to use nvprof (a general-purpose GPU
performance profiler).

\subsubsection{Qualitative Impressions}
\begin{table}
  \centering
  \footnotesize
  \begin{tabularx}{\columnwidth}{lXXcXX}
    \toprule
    & \multicolumn{2}{c}{\textbf{Useful}} & & \multicolumn{2}{c}{\textbf{Easy
        to Use}} \\
    \cmidrule{2-3}\cmidrule{5-6}
    \textbf{Feature} & Mean & Median & & Mean & Median \\
    \midrule
    Throughput Bar Chart & 4.4 & 4 & & 4.3 & 4 \\
    Memory Use Bar Chart & 4.4 & 4 & & 4.4 & 4 \\
    Run Time Breakdown & 4.4 & 4 & & 4.7 & 5 \\
    Memory Breakdown & 4.9 & 5 & & 4.6 & 5 \\
    Inline Markers & 4.0 & 4 & & 4.7 & 5 \\
    \bottomrule
  \end{tabularx}
  \caption{A summary of how useful and easy to use participants found
    \thetool{}'s features, on a five-point Likert scale (higher is better).}
  \label{tbl:qualitative-features}
  \vspace{-1.5em}
\end{table}

We first asked users to rate \thetool{}'s features based on whether they are
useful and easy to use; Table~\ref{tbl:qualitative-features} summarizes our
results. While the participants found all the features to be useful, these
results also seem to indicate that the memory breakdown was the most useful.
Additionally, the participants generally found each feature to be easy to use.

We then asked participants about the usefulness of the different interactions
offered by \thetool{}. These interactions were
\begin{enumerate*}[label=(\roman*)]
  \item highlighting the relevant line(s) of code when hovering over a
    visualization (mean 4.6; median 5),
  \item updating the visualizations when the code is changed (4.6; 5),
  \item double clicking breakdown bars to navigate the hierarchy (4.9; 5), and
  \item making batch size predictions using the draggable visualizations (3.9;
    4).
\end{enumerate*}
We also asked the participants whether they agreed that finding run time and
memory bottlenecks would be easy with \thetool{}. These two questions also
elicited positive responses, both receiving scores of 4.7 on average (median
5).

We believe that one factor that affected the participants' perceptions of the
draggable visualizations was that they were difficult to drag, due to an
implementation issue (the user's cursor needed to remain inside the bar chart
for any manipulations to take effect). Several participants ran into this issue
but were able to perform the dragging after being told to keep their cursor
inside the bar chart. We have since fixed this problem.

Finally, we asked the participants how strongly they agreed with two
statements about \thetool{} as a whole:
\begin{enumerate*}[label=(\roman*)]
  \item \textit{``By being present inside the text editor, Skyline could help
      me proactively develop my models with computational performance in
      mind''} (mean 4.1, median 4), and
  \item \textit{``Skyline's interactive features could help me better
      understand the computational training performance of my models''} (mean
    4.7, median 5).
\end{enumerate*}
We believe that these results are encouraging and pave the way for further
research toward improving performance understanding through interactivity.

Overall, the participants made positive comments throughout the study.
Participant~1 remarked that \textit{``[Skyline] is very intuitive''} and that
getting a memory breakdown \textit{``[...] is usually very hard to do.''}
Participant 2 mentioned that using \thetool{} was educational because they
learned about the run times and memory characteristics of various DNNs.
Furthermore, three out of the seven participants also voluntarily commented
that they would be interested in using \thetool{} in their own work.

\section{Discussion}\label{sec:discussion}
While the results of our study are promising, there are also limitations and
opportunities for additional research (and extensions of \thetool{}) that
warrant further discussion.

\subsection{Limitations and Future Work}
\namedparagraph{Model Quality.}
An ideal development tool for DNN training would bring together diagnostics and
predictions for both
\begin{enumerate*}[label=(\roman*)]
  \item computational performance, \emph{and}
  \item a model's prediction accuracy.
\end{enumerate*}
This set of features would empower deep learning developers to evaluate the
trade-offs between a model's computational performance and accuracy during
development. However, \thetool{} focuses only on the former as it does not
support making predictions about a DNN's potential prediction accuracy. This is
because predicting how a DNN's design (and its batch size) affects its accuracy
is a difficult problem; it boils down to understanding the factors that affect
a DNN's ability to ``learn,'' which is still an active area of research within
the machine learning community~\cite{lottery-frankle19,chamnet-dai18}. We see
\thetool{} as one step toward achieving this vision for ideal deep learning
development tools and we think that exploring ways to integrate \thetool{} with
prior work on deep learning accuracy
diagnostics~\cite{visinspect-krause16,seq2seqvis-strobelt19,activis-kahng18}
would be a great opportunity for future work.

\namedparagraph{User Study.}
Our user evaluation was an exploratory qualitative study that focused on
\thetool{}'s diagnostic capabilities. As a result, we do not make any claims
about
\begin{enumerate*}[label=(\roman*)]
  \item whether \thetool{} is a more effective performance debugging tool when
    compared to existing commercial profiling tools, nor
  \item how \thetool{}'s debugging capabilities might influence a user's DNN
    design process.
\end{enumerate*}
The goal of our study was to elicit early qualitative feedback from deep
learning developers about the effectiveness of in-editor, interactive
performance visualizations for DNN training. Our encouraging results indicate
that \thetool{}'s interactive features may be a promising way to communicate
complex performance information to deep learning developers. As described
above, we see \thetool{} as one step toward a comprehensive DNN training
development tool and we think that our results open up opportunities for
further research studies that can examine how having in-editor diagnostics for
both performance and accuracy might influence a user's DNN design process.

\subsection{Extensibility}
Although we implemented \thetool{} to work with Atom and PyTorch and to make
mutations to a DNN's batch size, the ideas behind \thetool{} are not
fundamentally limited to this editor, framework, and model parameter
combination.

\namedparagraph{Supporting Other Frameworks.}
To gather and organize performance data, \thetool{} relies on two concepts:
\begin{enumerate*}[label=(\roman*)]
  \item \emph{execution graphs} (a representation of the operations that run
    during a training iteration, along with their dependencies), and
  \item code ``modules.''
\end{enumerate*}
\thetool{} uses execution graphs to profile the individual operations in a
model and uses modules in the breakdown hierarchy. Both of these concepts exist
in other popular software frameworks. For example, both
TensorFlow~\cite{tensorflow-abadi16} and MXNet~\cite{mxnet-chen15} use graphs
to represent DNNs internally and both frameworks have the notion of ``modules''
(\texttt{tf.keras.Model} in TensorFlow and \texttt{gluon.nn.Block} in MXNet).

\namedparagraph{Supporting Other Editors.}
Since \thetool{} consists of an editor plugin and a daemon process that
communicate over a network socket, supporting additional editors is just a
matter of implementing a plugin for that specific editor. Any new editors just
need to provide APIs to allow the new \thetool{} plugin to
\begin{enumerate*}[label=(\roman*)]
  \item render user interface elements,
  \item open files, and
  \item mutate the contents of opened files.
\end{enumerate*}

\namedparagraph{Mutating Other Model Parameters.}
Supporting code mutations to other model parameters (e.g., the size of a layer)
would consist of three tasks:
\begin{enumerate*}[label=(\roman*)]
  \item developing performance models that relate the parameter to the model's
    training throughput and memory usage,
  \item capturing the relationship between inputs of successive layers (e.g.,
    changing the size of one layer may alter the size of its output, which in
    turn affects the input size of the next layer in the DNN), and
  \item identifying how the parameter is specified in the code.
\end{enumerate*}
Task (i) can be approached using regression methods similar to \thetool{}'s
batch size performance models. Task (ii) is a matter of implementing the
well-defined mathematical rules that relate a DNN layer's size to its expected
input and output sizes. Task (iii) can be approached using the code's abstract
syntax tree because the way DNN parameters are specified is ultimately imposed
by the software framework (e.g., PyTorch).

\subsection{Additional Use Cases}
In addition to being a performance profiling and debugging tool, we believe
that \thetool{} has the potential to help users in other ways.

\namedparagraph{Proactive Model Design.}
\thetool{} shortens the performance feedback loop by displaying computational
performance information live during development. We think that this rapid
feedback could enable deep learning developers to be more \emph{proactive} when
designing their models for performance because they would be able to quickly
experiment with and iterate on model designs during development.

\namedparagraph{Education.}
\thetool{}'s interactive features could also be used to help teach deep
learning developers about the performance characteristics of DNN training. For
example, users can drag the throughput bar chart to see the relationship
between batch size, throughput, and memory usage (e.g., throughput tends to
increase with larger batch sizes, but with diminishing returns). They can also
use \thetool{}'s breakdowns to see the performance costs of different
operations instead of guessing about performance---a common approach used by
our interviewees.

\section{Conclusion}
This paper presents \thetool{}: a new interactive tool for DNN training that
supports in-editor performance profiling, visualization, and debugging.
The key idea behind \thetool{} is that it leverages special properties of DNN
training (repetitiveness, predictability, structured code) to offer interactive
performance visualizations tied to the code.
The main takeaways from this work are that
\begin{enumerate*}[label=(\roman*)]
  \item DNN training computational performance debugging is an important
    problem faced by deep learning developers, and
  \item DNN training computation has useful \emph{properties} that can enable
    new interactive features (e.g., directly manipulatable visualizations that
    mutate the batch size in the code).
\end{enumerate*}
Finally, we have open-sourced \thetool{} at
\href{https://github.com/skylineprof/skyline}{github.com/skylineprof/skyline}
to help deep learning developers and to hopefully facilitate future research in
this area.

\section{Acknowledgments}
We thank the anonymous reviewers for their feedback. We also thank the
participants in our formative and user studies for providing their insights and
time. We are grateful to James Gleeson and Hongyu Zhu for providing feedback on
earlier versions of the \thetool{} project, as well as to all members of the
EcoSystem research group for the stimulating research environment they provide.
This work was supported by a Queen Elizabeth II Graduate Scholarship in Science
and Technology, Vector Scholarship in Artificial Intelligence, Snap Research
Scholarship, and an NSERC Canada Graduate Scholarship -- Master's (CGS M). This
work was also supported in part by the NSERC Discovery grant, the Canada
Foundation for Innovation JELF grant, the Connaught Fund, and Huawei grants.

\balance{}

\bibliographystyle{SIGCHI-Reference-Format}
\bibliography{references/deep-learning,references/ml-systems,references/urls,references/hci}

\end{document}